\documentclass[namedreferences]{solarphysics}
\usepackage[optionalrh,solaromanenum]{spr-sola-addons} 
\usepackage{graphicx}        
\usepackage{color}           
\usepackage{url}             

\begin{document}
\begin{article}
\begin{opening}
\title{%
Resolving Azimuth Ambiguity Using Vertical Nature of Solar Quiet-Sun Magnetic Fields
}
\author{S.~\surname{Gosain}\sep
        A.A.~\surname{Pevtsov}
}
\runningauthor{Gosain and Pevtsov}
\runningtitle{Resolving Azimuth Ambiguity of Quiet-Sun Magnetic Fields}
\institute{ National Solar Observatory, 950 N. Cherry Avenue, Tucson 85719, Arizona, USA
	email: \url{sgosain@nso.edu} \\
}
\begin{abstract}
The measurement of solar magnetic fields using the Zeeman effect diagnostics has a fundamental $180^\circ$ ambiguity in the determination of the azimuth angle of the transverse field component. There are several methods that are used in the community and each one has its merits and demerits. Here we present a disambiguation idea that is based on the assumption that most of the magnetic field on the sun is predominantly vertical. While the method is not applicable to penumbra or other features harboring predominantly horizontal fields like the sheared neutral lines, it is useful for regions where fields are predominantly vertical like network and plage areas. The method is tested with the full-disk solar vector magnetograms observed by the VSM/SOLIS instrument. We find that statistically about 60--85 \% of the pixels in a typical full-disk magnetogram has field inclination in the range of $0^\circ$--$30^\circ$ with respect to the local solar normal, and thus can be successfully disambiguated by the proposed method. Due to its non-iterative nature, the present method is extremely fast and therefore can be used as a good initial guess for iterative schemes like nonpotential field computation (NPFC) method. Furthermore, the method is insensitive to noisy pixels as it does not depend upon the neighboring pixels or derivatives.
\end{abstract}
\keywords{Active Regions, Magnetic fields, Polarimetry}
\end{opening}

\section{Introduction}
\label{S-Introduction}
The measurement of solar magnetic field vector is very important for our understanding of the solar activity phenomena like flares and CMEs. These explosive solar events affect the geospace, thereby impacting our space endeavors. The free energy that is available to fuel these explosive events is mostly stored in the non-potential component of the solar magnetic field. The study of the development of non-potentiality in active region magnetic fields requires high-quality vector magnetogram observations, ideally at a good cadence. The rate at which the solar vector magnetograms are being obtained by various ground and space-based instruments has therefore increased dramatically, especially during the present decade. Some of the observatories that are routinely providing the vector magnetograms, among others, are Spectro-Polarimeter (SP) on {\it Hinode}/Solar Optical Telescope (SOT) (Tsuneta {\it et al.}, 2008) from space, over a limited field-of-view (FOV), and the vector spectro-magnetograph (VSM) on the Synoptic Optical Long-term Investigation of the Sun (SOLIS) facility (Keller {\it et al.}, 2003; Jones {\it et al.}, 2005; Henney {\it et al.}, 2006) from ground, over the full-disk. Both of these instruments are slit-scanning type spectro-polarimeters. The Helioseismic and Magnetic Imager (HMI) on {\it Solar Dynamics Observatory} (SDO; Pesnell {\it et al.}, 2012) is a two-dimensional imaging spectro-polarimeter, employing a tunable narrow band filter. Here we disregard specific differences between two types of magnetographs, and concentrate on their common problem - a resolution of 180$^\circ$ ambiguity of azimuths of transverse fields.

All of the magnetic field measurements making use of the Zeeman effect diagnostics have intrinsic $180^\circ$ azimuth ambiguity, {\it i.e.} it does not distinguish between the two possible azimuth directions, $\phi$ and $\phi$+180$^\circ$, of the transverse field component. A number of ambiguity resolving algorithms have been developed which are discussed and compared in detail in a review by Metcalf {\it et al.} (2006). The two algorithms that scored very well in that comparative study, {\it i.e.}, minimum energy (ME) and nonlinear minimum energy (NME) methods require significant processing time, which may be a serious issue when the disambiguation needs to be done in a near real-time for full-disk data with large number of pixels.

With the modern solar instrumentation the number of pixels  to be solved for ambiguity is increasing: For example, {\it Hinode}/SP maps of active regions are already of substantial size ($\lesssim$ a million pixel) and future Advanced Technology Solar Telescope (ATST) data will be even larger.
Further, the full-disk observations need to resolve ambiguity in a large number of pixels (typically 4 and 16 million pixels for SOLIS/VSM and SDO/HMI vector magnetograms, respectively). This speed issue is thus a major factor in the selection of disambiguation algorithm for full-disk vector magnetograms.

In the case of SOLIS/VSM full-disk vector magnetograms, which are used in the present work, the nonpotential field computation (NPFC) algorithm (Georgoulis, 2005; Georgoulis {\it et al.}, 2008) has been implemented mainly due to its overall successful disambiguation and its speed. In practice, it takes 2--3 h to disambiguate a typical full-disk SOLIS/VSM $2048 \times 2048$ pixel magnetogram.

While efforts are on-going to develop more efficient and faster algorithms, here we present a simple disambiguation method which can serve as a first order disambiguation algorithm to speed up more robust disambiguation schemes, by providing a good initial guess for most of the pixels and thereby facilitating a faster convergence to the minima of the merit function. The algorithm is based on the observational fact that most of the magnetic field outside the sunspots or sheared polarity inversion lines (PIL) are found to be nearly vertical at the photospheric level. With this fact in mind we can then resolve the azimuth ambiguity in the image plane within certain constraints. We shall first illustrate the idea behind the present method and then use a full-disk vector magnetogram observation from SOLIS/VSM instrument to demonstrate that a major portion of the magnetic field on the solar surface is vertical, and therefore the current method is applicable for many pixels. We apply the present disambiguation method to the non-disambiguated SOLIS/VSM data and make comparison with the results obtained with the NPFC method. The merits, demerits and the constraints are discussed.

\subsection{Description of the Method}
The present method can be illustrated by a schematic shown in the top panel of Figure 1. A simple scenario is sketched where a magnetic field vector, represented by green arrow, is located at a heliocentric angle $\theta$. The dashed line shows the local solar vertical direction. Under the assumption that the magnetic field is close to vertical (henceforth called as the vertical field approximation or VFA), it may be noticed that the azimuth angle on the image plane would point radially outward ({\it i.e.}, toward the solar limb) for an apparent (line-of-sight [LOS]) positive polarity, and radially inward ({\it i.e.}, toward the solar disk center) for a LOS negative polarity. Following the convention that $[\theta \in (0 , \pi/2)]$ ({\it i.e.} symmetric treatment of eastern and western solar hemispheres) and $[\gamma \in (-\pi/2, \pi/2)]$, if the following two conditions are met then the above mentioned VFA method for disambiguation is applicable in general
$$\theta > |\gamma| ,$$
$$\theta + |\gamma| < \pi/2 .$$

In other words, as long as the field is vertical to the extent that its inclination angle $|\gamma|$ with respect to the local solar vertical is less than the heliocentric angle $\theta$. This can be visualized as follows. Consider positive polarity patch as shown in the lower left panel of Figure 1. For all possible azimuth directions (loci shown by red ellipse) the solution in the image plane is the one which makes an acute angle with respect to the radius vector, CR. Similarly, for a negative polarity patch, as shown in the lower right panel of Figure 1, one would choose azimuth solution which makes an obtuse angle with respect to the radius vector, CR. One can notice that this method works better for larger heliocentric angles where the VFA, $\theta > |\gamma|$, is more likely to be satisfied for predominantly vertical fields. As one approaches closer to the disk center the condition of VFA would be fulfilled by the fields which are very close to vertical.

The second condition becomes important closer to the limb, {\it i.e.}, when the local solar vertical is closer to the plane-of-sky and so $\theta +|\gamma|$ can exceed $\pi/2$ for non-vertical fields simply because $\theta$ itself is nearing $\pi/2$ close to the limb. When $\theta +|\gamma|$ exceeds $\pi/2$, the VFA method fails as the LOS polarity of field in the image plane flips its sign, {\it i.e.}, a positive polarity field would appear negative. Such a condition also happens if the region is not close to the limb but the field vector is significantly horizontal, for example in sunspot penumbrae. The red ellipse shown in Figure 1 can span on either side of the line-of-sight as well as beyond the plane-of-sky, hence the azimuth in the image plane cannot be resolved using the VFA method.

In the following sections we will use the real solar observations of the magnetic field vector over the full-disk obtained by the SOLIS/VSM instrument to show that a large fraction of the magnetic field at the photosphere is close to vertical and therefore can be resolved using the VFA method. It should be noted however that the VFA disambiguation only provides a good initial guess for azimuth solution and a more elaborate scheme like the NPFC method should be applied as second step to take advantage of this good guess.

A technically different approach to perform disambiguation based on inclination of the field would be, to deduce two possible inclination angles of the field at each pixel in the heliographic system corresponding to the two possible azimuth solutions in the image plane and choose the one that gives more vertical field. However, in this approach we would need to (a) consider both possible azimuth solutions in the image plane, (b) for each possible solution, transform the vectors in the heliographic system and then evaluate the inclination angle ($|\gamma|$) with respect to the local solar vertical, and (c) finally choose the solution which gives more vertical field. This requires an extra step of heliographic transformation of vectors at each pixel and needs information of $L$, $B$, $L_0$, $B_0$, and $P$-angle for transformation. On the contrary, the VFA method requires neither this information nor heliographic transformation. This makes the VFA method much faster and less expensive computationally.

\subsection{The Distribution of Inclination Angle at the Photosphere: SOLIS/VSM Observations}
The assumption regarding the field at the photosphere being generally mostly vertical can be verified from the observations themselves. To evaluate the VFA applicability we use a full-disk vector magnetogram observed by SOLIS/VSM on 14 August 2011. There were no major sunspot groups present on the day of observations, but still there was enough flux uniformly distributed on the solar surface in network and facular areas. Since the VFA method is applicable to such flux concentrations and not to sunspot penumbrae and PIL with horizontal fields, we chose this magnetogram as a representative case. The magnetic field vector is derived by performing the inversion of the full Stokes profiles of the solar disk observed in the Fe {\sc i} 630 nm line pair in the framework of the Milne-Eddington (ME) model of stellar atmospheres and following the approach described by Lites and Skumanich (1990). Only the pixels with polarization signal greater than 0.5\% of continuum intensity are inverted in order to avoid noise. Additional details about the SOLIS/VSM instrument and the calibration of its polarization can be found in Keller {\it et al.} (1998, 2003) and in Jones {\it et al.} (2005). The field azimuth is disambiguated by using the NPFC method (Georgoulis, 2005). The vector field is then transformed into heliographic coordinates using the method of Gary and Hagyard (1990), to get the true radial and tangential components of the magnetic flux. Using these transformed vector field components we get the true field inclination on the solar surface. A map of radial flux is displayed in the top panel of Figure 2 with a scaling of $\pm$ 250 G (gauss). The distribution of the inclination angle is shown in the lower panel of figure 2. The lower left panel shows the relative frequency of the inclination angle while the lower right panel shows the cumulative histogram of the inclination angles. It should be noted that, for the purpose of the test, we have converted the range of field inclination angles from $0^\circ$--$180^\circ$ to $0^\circ$--$90^\circ$ ({\it i.e.}, ignoring the polarity of the field) as we want to check the VFA assumption irrespective of the polarity of the field. From the histogram we see that most of the field in this full-disk magnetogram is close to vertical. From the cumulative histogram, the field which is inclined in the range $0^\circ$--$30^\circ$ comprises of about 85\% of the pixels. A similar study for other dates shows that a typical fraction of the field which is inclined in the range of $0^\circ$--$30^\circ$ is $\approx$60--85\% of all (full-disk magnetogram) pixels. Thus, the basic underlying assumption of VFA is valid for a large portion of the solar disk.

\subsection{Distribution of Field Azimuth with Respect to Radial Direction}
One may note that for a purely vertical positive polarity field distributed on the solar disk the 180$^\circ$ ambiguous, transverse field azimuths on the image plane would exactly point in the radial direction away from the disk center. In Figure 3 we show the full-disk longitudinal magnetogram overlaid with blue and red line segments which represent, respectively, the general direction of the non-disambiguated transverse field component in the image plane of (a) the actual observed field, and (b) the general direction if the field were purely vertical. If the observed field were purely vertical, {\it i.e.} $\gamma$=0$^\circ$ or 180$^\circ$, then we would expect the red and blue line segments to overlap each other at all locations. However, due to a distribution of inclination angles shown in Figure 2 (lower panels), this is not the case. A small deviation from the pure verticality and a random orientation of the field azimuth (the red ellipse in Figure 1) lead to a range of angular deviations between the radial direction and the field azimuth in the image plane. It may be noted that we can solve the azimuth ambiguity based on the observed LOS polarity of the field not just for vertical field but a range of inclination angles which satisfy the criterion described in Section 1.1.

\subsection{Testing the VFA Method with Respect to NPFC Method}
We now apply the VFA disambiguation method to solve the azimuth ambiguity of the SOLIS/VSM observations during 14 August 2011. For comparison we use the NPFC disambiguated SOLIS/VSM magnetogram as a reference. The NPFC disambiguated azimuth map for this magnetogram is shown in Figure 4 in $0^\circ$--$360^\circ$ color map. The basic idea behind the NPFC algorithm (Georgoulis {\it et al.}, 2008) is that any magnetic field {\bf B} can be decomposed into current-free (potential) component {\bf B}$_{\rm p}$ and a current-carrying component {\bf B}$_{\rm c}$, {\it i.e.}, {\bf B}={\bf B}$_{\rm p} +$ {\bf B}$_{\rm c}$. By using the vertical component (normal to the photospheric boundary, say $S$) of the field, $B_z$, and current density, $J_z$, one can compute {\bf B}$_{\rm p}$ and {\bf B}$_{\rm c}$ under the assumption that $(\partial B_{{\rm c}z}/\partial z)|_S=0$. The problem that both $B_z$ and $J_z$ required to compute {\bf B}$_{\rm p}$ and {\bf B}$_{\rm c}$ depend on the azimuth solution themselves is then solved iteratively. A good initial guess, for example using VFA, would therefore prove to be useful for improving the performance of NPFC. The full-disk disambiguation is performed by dividing the full disk into smaller tiles and applying NPFC successively to them. A typical time required for the full-disk disambiguation using the NPFC method is about 2--3 h. However, using the VFA method the solution is instantaneous and takes only a few seconds. The azimuth solution found using VFA for the 14 August 2011 observation is compared to the NPFC result by taking a difference between the two azimuth solutions. The result is displayed in Figure 5, where only the residuals which are non zero can be seen in black-white shades while rest of the background gray shades correspond to zeroes. It may be noted that as one approaches closer to the disk center the VFA conditions (1) and (2) start to fail as the peak distribution of $\gamma\approx$$0^\circ$--$30^\circ$ is no longer less than the heliocentric angle $\theta$ and so closer to disk center we find that the VFA method does not solve the ambiguity. However, the fraction of pixels for which the azimuth solutions by the two methods agree is about 85\%. For typically randomly chosen magnetograms by SOLIS/VSM on other dates this range of agreement was found to be in the range of 60--85\%, depending on the distribution of magnetic flux on the solar surface. This agreement of 60--85\% essentially arises due to the fact that the VFA assumption is valid for 60--85\% of the full-disk pixels. If all the pixels on a full disk were to follow VFA assumption, then the match between the VFA and NPFC disambiguation would also approach 100 \%. Therefore, in that sense the success rate of the VFA method is 100 \%, and the correct disambiguation is only limited by the number of pixels actually following the VFA assumption. When sunspots dominate the surface flux the VFA assumption is poorer for larger portion of solar disk and therefore the performance of VFA is expected to be lower.

\section{Discussion and Conclusions}
The vertical nature of magnetic field has been investigated earlier by several authors using a variety of methods. Svalgaard {\it et al.} (1978) found that the center-to-limb variation of average line-of-sight field strength within an equatorial band follows the cos($\rho$) function (where $\rho$ is the heliocentric angle) which one expects for radial fields. Similar results were found by Howard (1974) by comparing the magnetic flux of a region located at equidistant longitudes east/west of the central meridian. He concluded that the east-west tilt component of the photospheric field was generally less than 10$^\circ$. Using Wilcox Observatory data, Shrauner and Scherrer (1994) performed a least-square fit to a simple projection model and found similar values of tilts. More recently, using five years worth of SOLIS/VSM full disk longitudinal magnetograms, Petrie and Patrikeeva (2009) showed that the east-west inclination of most of the photospheric field is less than about 12$^\circ$.

Here, we showed using the SOLIS/VSM full-disk solar vector magnetograms that 60--85\% of the field on the solar surface is vertical to within $0^\circ$--$30^\circ$ range. We then showed how most of these pixels satisfying the VFA approximation can be solved for azimuth ambiguity relatively quickly and easily using the procedure described in this paper. The results of the method agree with those from the NPFC method with a success rate of 60--85\% depending on flux distribution. The main merit of the method is its relative simplicity and speed of the method which it can be exploited in many ways to speed up the disambiguation of the vector magnetograms, particularly full-disk observations where the number of pixels to be disambiguated tend to be very large. However, the method is not stand alone and needs to be used, as a good initial guess, along with more robust schemes like NPFC in order to disambiguate the full-disk data completely. An advantage of the VFA method is that it is applicable even close to the solar limb. 
 Further, as VFA is based on single pixel information and does not involve derivatives it is not affected by the effects of seeing noise or non-uniformity in mapping due to slit scanning. Such quick disambiguation algorithms are required to provide good initial guesses for real-time ambiguity resolution of contemporary vector field measurements, for example, SDO/HMI which observes vector fields at 16 million pixels format at a cadence of 12 min.

Further, the present VFA method needs to be discussed in the context of recent finding of the so-called ``seething" horizontal magnetic fields in the quiet Sun by Harvey {\it et al.} (2007). It was found that the photospheric magnetic field outside of active regions and the network exhibits a ubiquitous and dynamic line-of-sight component which is stronger towards the limb. This observation, interpreted as horizontal field component, shows a time variation with an average temporal rms near the limb of 1.7 G at $\approx 3''$ resolution. SOLIS/VSM full-disk vector magnetograms are observed at a resolution of $\approx 2''$ (as determined by typical seeing and $1''$ pixel sampling). However, the Milne-Eddington inversion of the Stokes profiles is done only when the signal-to-noise ratio of the polarization signal is above a certain threshold. This threshold is taken as $\geq 1\times10^{-3}$ of the continuum intensity, and limits the observed field strengths to a few tens of gauss, whereas the horizontal fields in quiet sun field detected by Harvey {\it et al.} (2007) were found for fields less than 5.5 G. So, the VFA method applied to the SOLIS/VSM magnetograms or magnetograms from other instruments with similar resolution and/or noise characteristics would not be affected by the horizontal fields as they are below the observed values.

\begin{figure}    
\centerline{\includegraphics[width=1.0\textwidth,clip=]{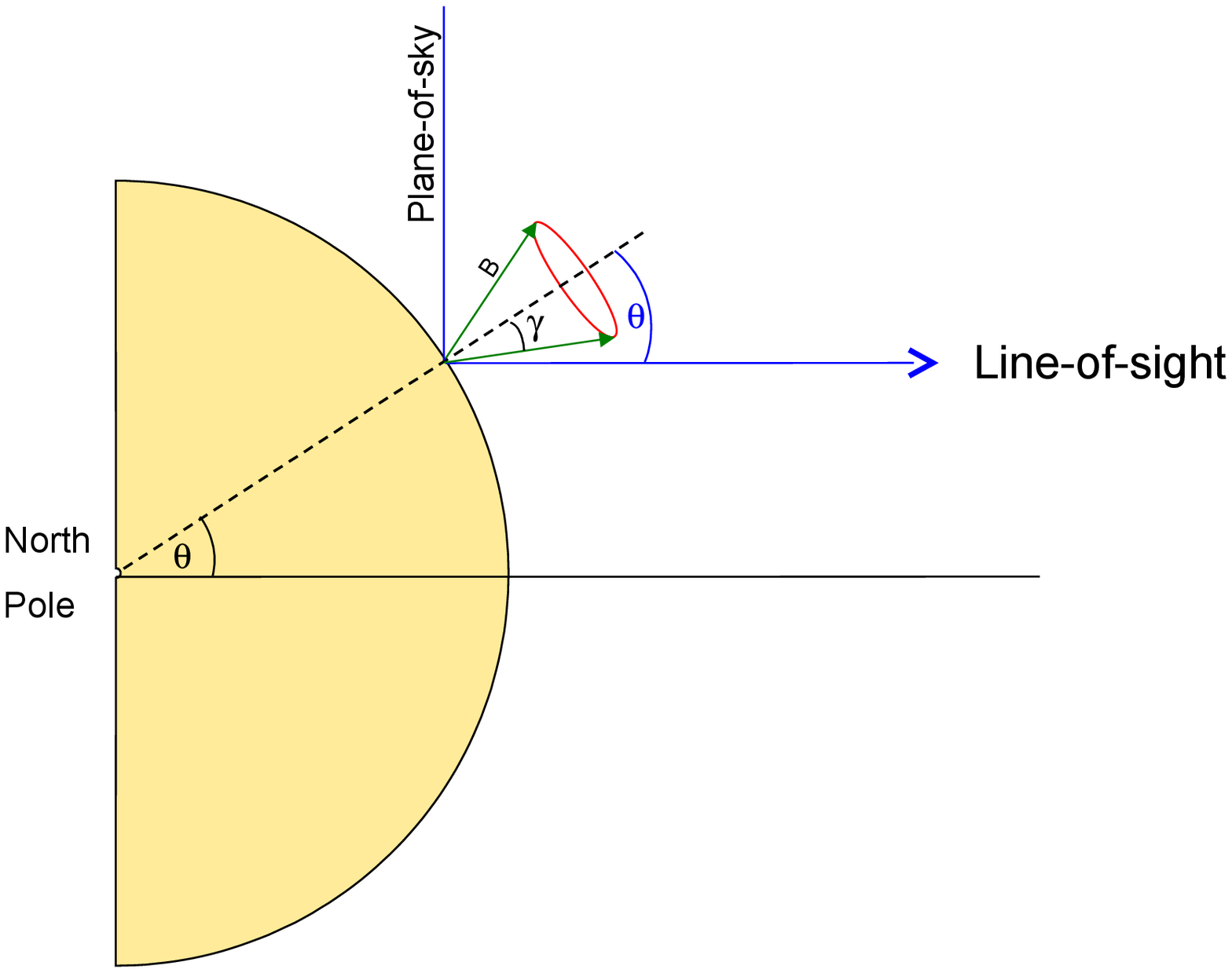}}
\centerline{\includegraphics[width=1.0\textwidth,clip=]{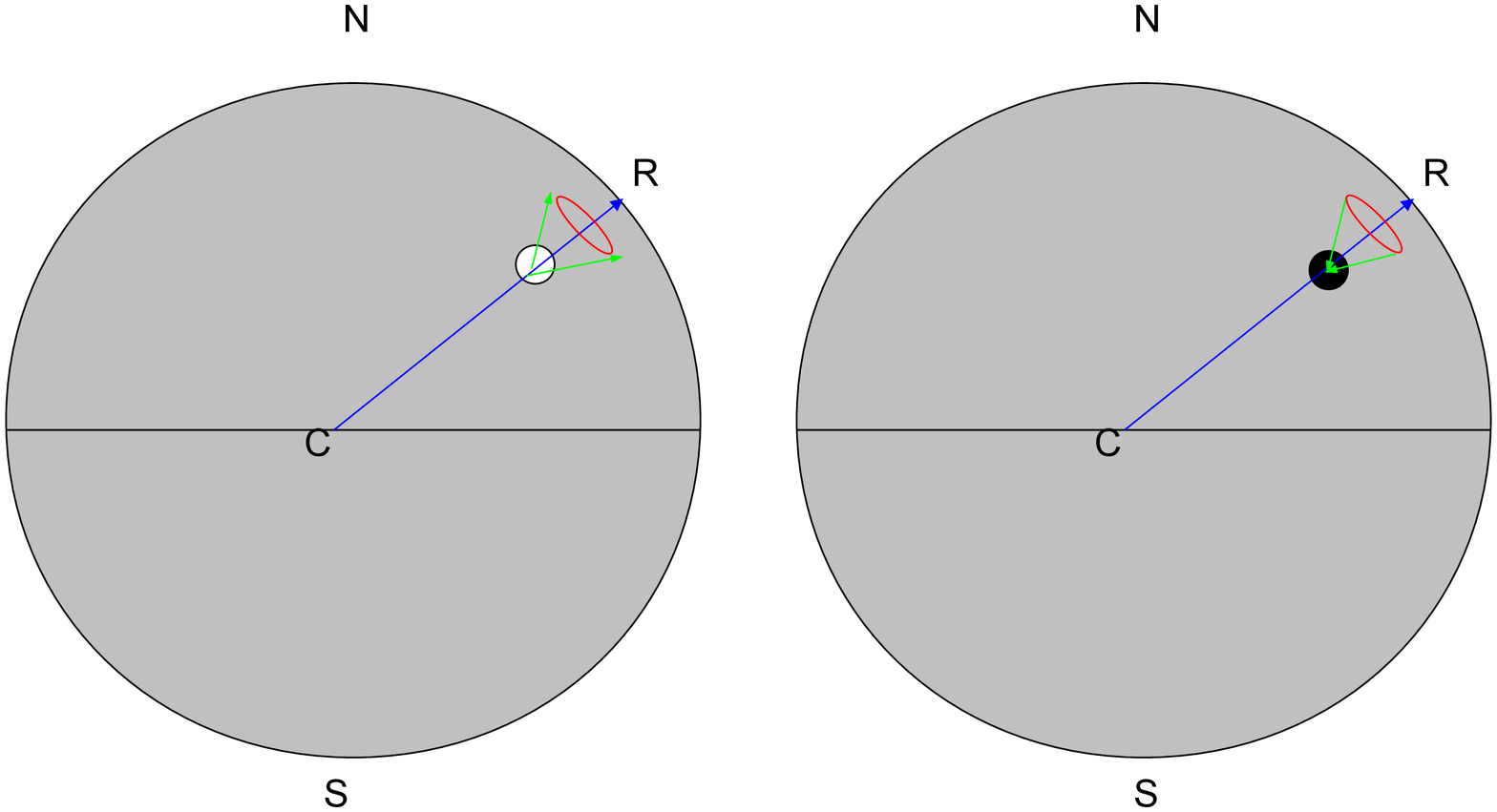}}
\caption{%
The top panel illustrates the idea behind the VFA ambiguity resolution method. A portion of the solar disk as viewed from the solar north pole is shown. A typical magnetic field vector {\bf B} (represented by green arrow) is assumed to be close to vertical making an inclination angle $\gamma$ with respect to the local solar normal. The field vector {\bf B} is assumed to be located at a heliocentric angle $\theta$. Blue arrow points towards the line-of-sight direction and the plane-of-sky is represented by the vertical blue line. The ellipse (shown in red color) represents the loci of all the possible azimuth directions in which the field vector {\bf B} with inclination $\gamma$ can lie. As described in Section 1.1, the azimuth direction in the image plane can be correctly resolved using the VFA method only when $\theta > |\gamma|$ and $\theta +|\gamma| <\pi/2$. The lower panel illustrates the two cases when the LOS component of the field over a patch is positive (white circular patch in the lower left panel) and negative (black circular patch in the lower right panel). The azimuth solution over the observed patch in the image plane is simply the direction which makes an acute (obtuse) angle with the radius vector CR for positive (negative) polarity of the LOS field, respectively.
}
\label{figure1}
\end{figure}

\begin{figure}   
\centerline{\includegraphics[width=1.1\textwidth,clip=]{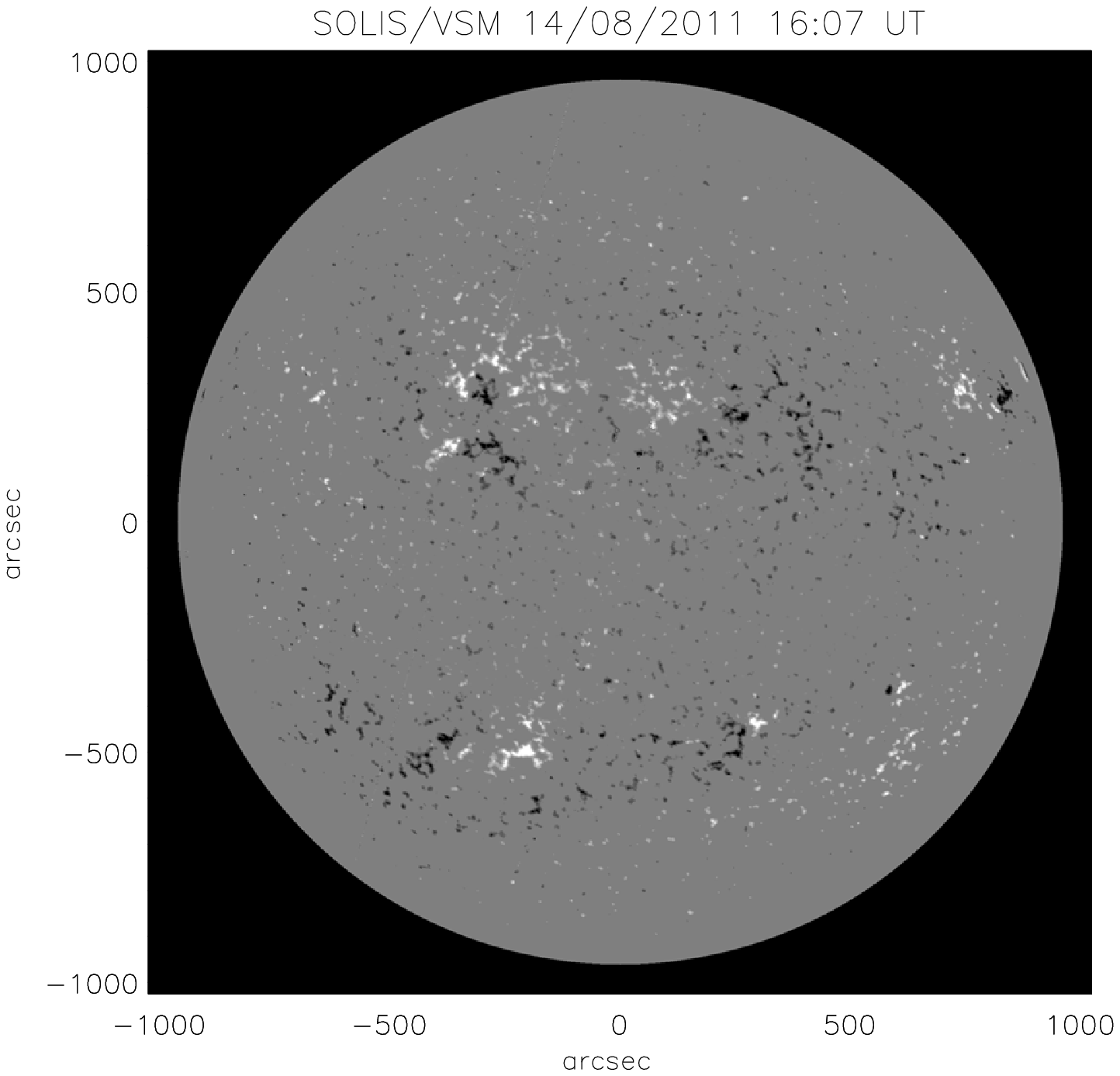}}
\centerline{\includegraphics[width=1.1\textwidth,clip=]{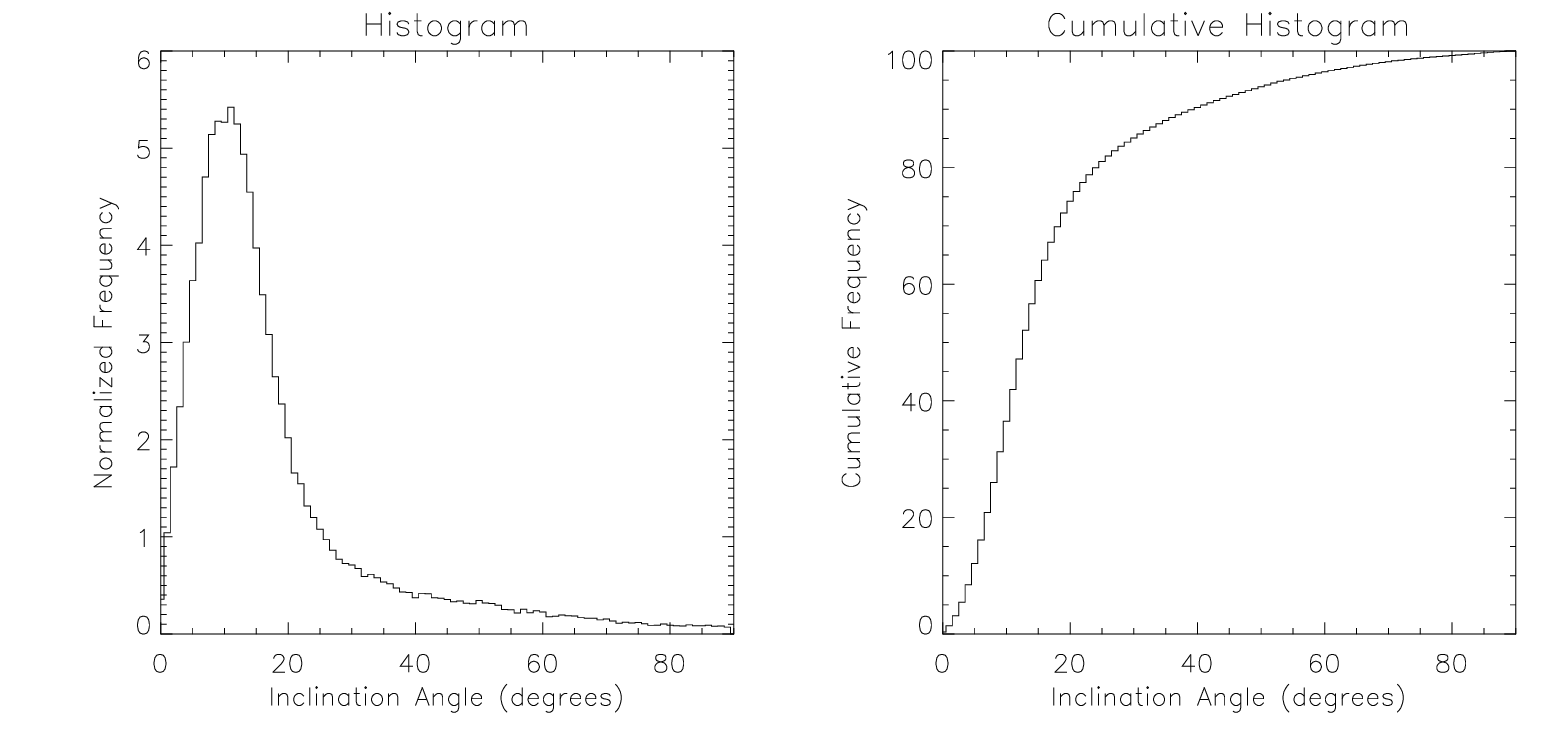}
}
\caption{%
The top panel shows the full-disk map of the radial component of the magnetic field vector derived from the SOLIS/VSM vector magnetogram after (i) disambiguation using the NPFC method, and (ii) transformation of the vectors to the heliographic coordinates. The lower left panel shows the histogram of the inclination angle $\gamma$ for all the pixels above polarization threshold of 0.5 $\%$ of continuum intensity. The lower right panel shows the cumulative histogram.
}
\label{figure2}
\end{figure}

\begin{figure}    
\centerline{\includegraphics[width=1.1\textwidth,clip=]{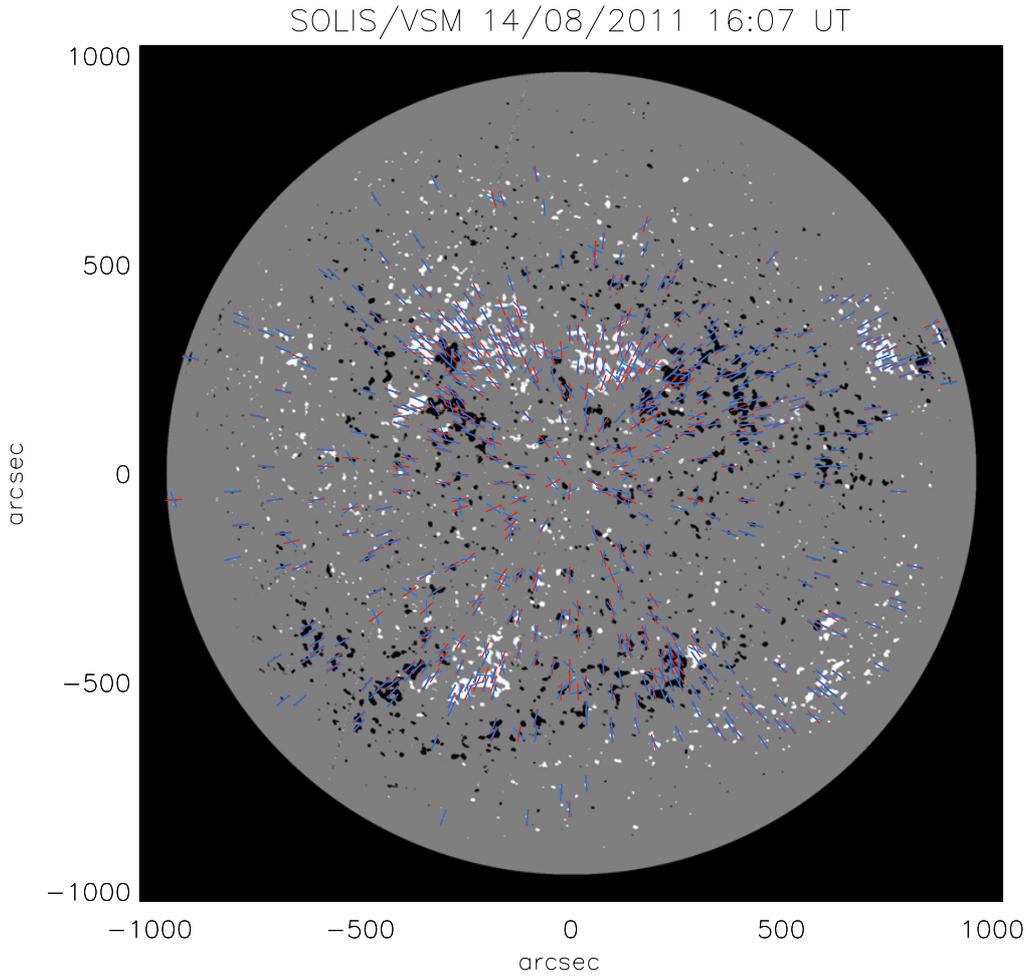}
}
\caption{%
The full-disk map of the vertical component of the magnetic field obtained after the NPFC disambiguation and heliographic transformation of SOLIS/VSM vector magnetogram. The red line segments represent the radial direction and the blue line segment shows the unresolved azimuth direction. It can be seen that most of the azimuths are closely aligned with the radial direction. The deviation increases closer to the disk center. For some locations the two line segments are exactly above one another such that the line segment of only blue color (which is on the top) can be seen.
}
\label{figure3}
\end{figure}

\begin{figure}    
\centerline{\includegraphics[width=1.0\textwidth,clip=]{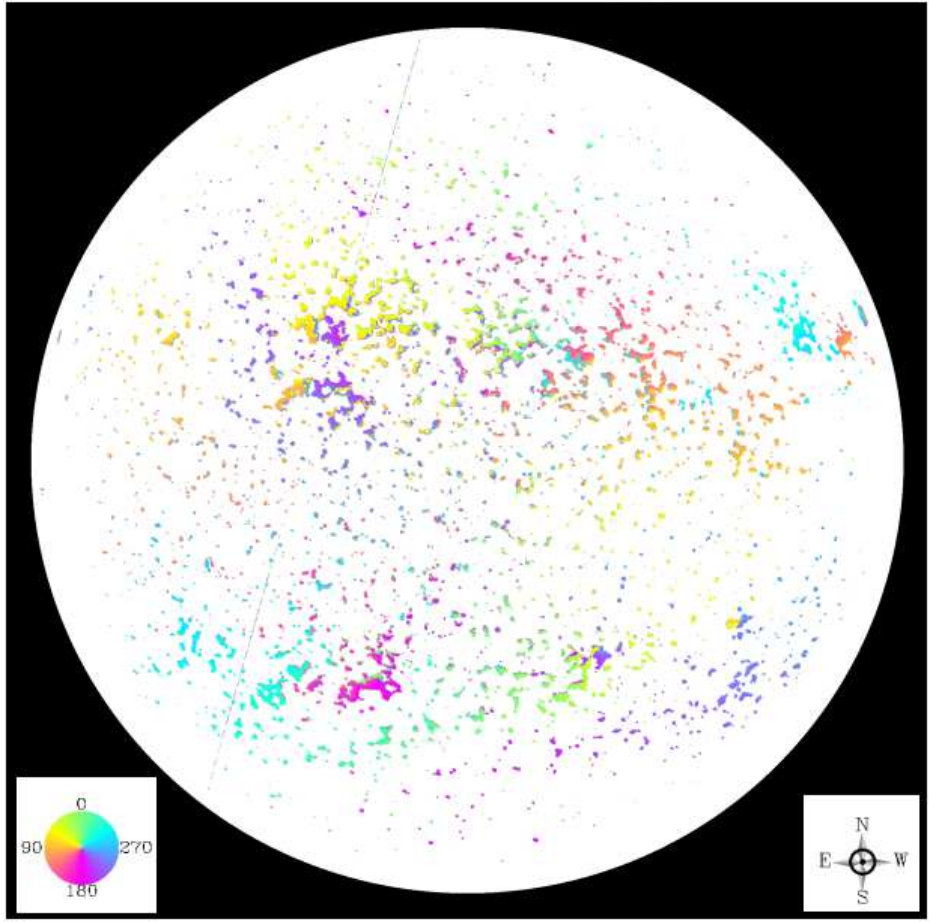}
}
\caption{%
A full-disk map of the field azimuths for 14 August 2011, after the NPFC disambiguation algorithm by the SOLIS/VSM pipeline code. The azimuth is measured counterclockwise from the solar north. A slanted stripe across the disk is due to a glitch in the camera during the scan.
}
\label{figure4}
\end{figure}

\begin{figure}    
\centerline{\includegraphics[width=1.0\textwidth,clip=]{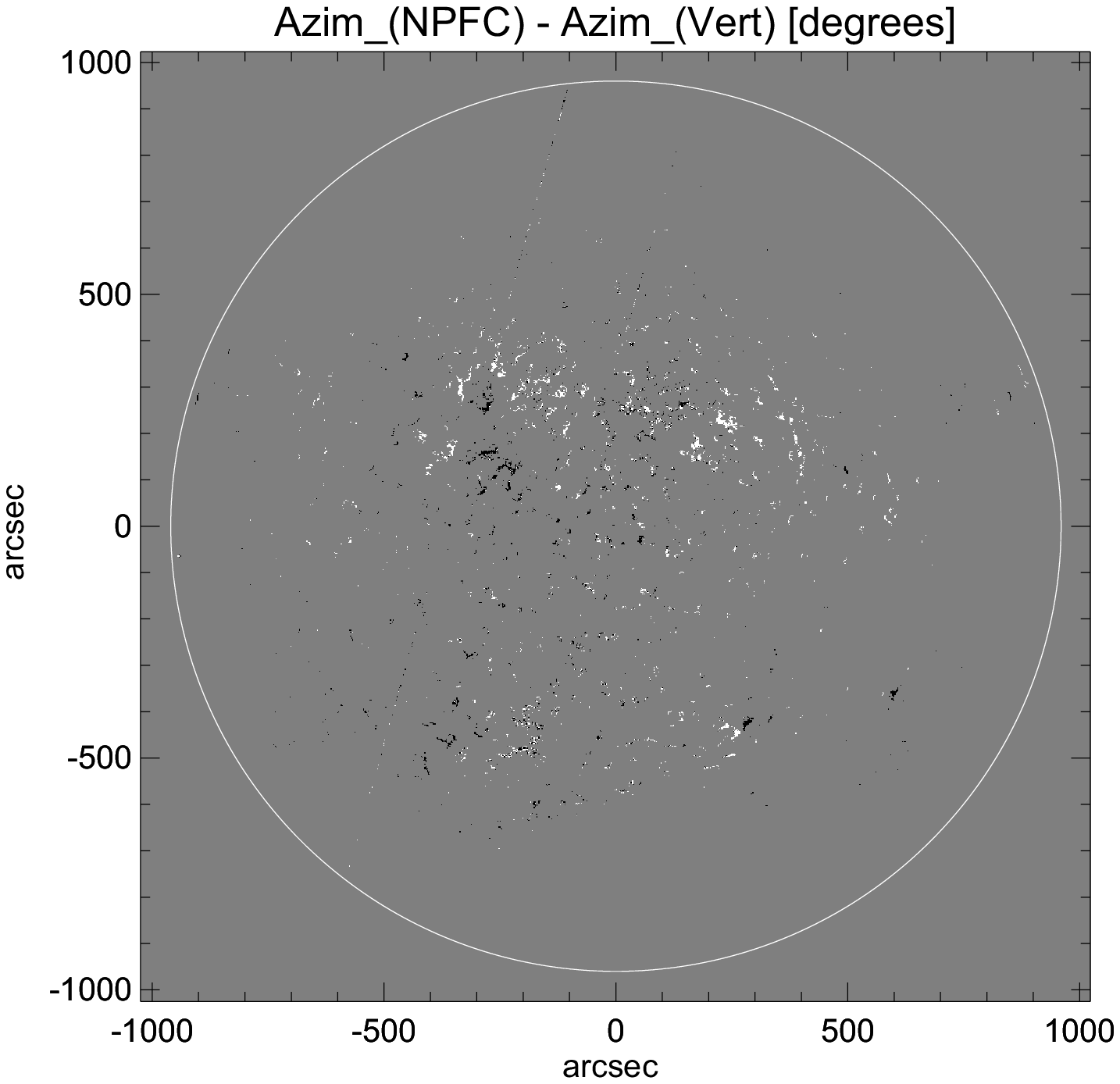}
}
\caption{%
The difference between the full-disk azimuth maps disambiguated by the present VFA method and the NPFC method for 14 August 2011 (full map after the NPFC disambiguation is shown in Figure 4). The locations where the ambiguity solution using the VFA approach does not match the NPFC solution (matching solution means the difference is 0$^\circ$, shown by grey shade, while mismatch of solutions means difference is $\pm180^\circ$, shown by white/black shade) can be seen to be more concentrated near the disk center where the applicability of VFA is limited.
}
\label{figure5}

\end{figure}

\begin{acks}
The authors thank the anonymous referee for providing critical comments which improved the presentation of the method in the paper. The authors also thank the SOLIS team for providing the full disk disambiguated magnetograms used in this study for testing the results of VFA approximation. SOLIS data used here are produced cooperatively by NSF/NSO and NASA/LWS. National Solar Observatory (NSO) is operated by Association of Universities for Research in Astronomy (AURA, Inc) under a cooperative agreement with the NSF.
\end{acks}

\bibliographystyle{spr-mp-sola}

\begin{thebibliography}{}
\bibitem[Gary and Hagyard(1990)]{1990SoPh..126...21G}Gary, G.A.,
Hagyard, M.J.: 1990, {\it Solar Phys.} {\bf 126}, 21.
\bibitem[Georgoulis(2005)]{2005ApJ...629L..69G}Georgoulis, M.K.: 2005, {\it
Astrophys. J. Lett.} {\bf 629}, L69.
\bibitem[Georgoulis, Raouafi, and Henney(2008)]{2008ASPC..383..107G}Georgoulis, M.K., Raouafi, N.-E.,
Henney, C.J.: 2008, In: Howe, R., Komm, R.W., Balasubramaniam, K.S., Petrie, G.J.D. (eds.), {\it Subsurface and Atmospheric Influences on Solar
Activity, ASP Conf. Ser.} {\bf 383}, 107.
\bibitem[Henney, Keller and Harvey(2006)]{2006ASPC..358...92H}Henney, C.J., Keller, C.U., Harvey, J.W.: 2006, In: Casini, R., Lites, B.W. (eds.),
{\it Solar Polarization 4, ASP Conf. Ser.} {\bf 358}, 92.
\bibitem[Howard(1974)]{1974SoPh...39..275H}Howard, R.: 1974, {\it Solar
Phys.} {\bf 39}, 275.
\bibitem[Jones \emph{et al.}(2005)]{2005AGUSMSP51A..02J}Jones, H.P.,
Malanushenko, O.V., Harvey, J.W., Henney, C.J., Keller, C.U.: 2005,
 AGU Spring Meeting, SP51A-02.
\bibitem[Keller(1998)]{1998SPIE.3352..732K}Keller, C.U.: 1998, In: Stepp, L.M. (ed.), {\it Advanced Technology Optical/IR Telescopes VI, Proc. SPIE} {\bf
3352}, 732.
\bibitem[Keller et al.(2003)]{2003SPIE.4853..194K} Keller, C.~U., Harvey,
J.~W.,  Giampapa, M.~S.:\ 2003, In: Keil, S.L., Avakyan, S.V. (eds.), {\it Innovative Telescopes and Instrumentation for Solar Astrophysics, Proc. SPIE} {\bf 4853}, 194.
\bibitem[Lites and Skumanich(1990)]{1990ApJ...348..747L}Lites, B.W.
Skumanich, A.: 1990, {\it Astrophys. J.} {\bf 348}, 747.
\bibitem[Metcalf(1994)]{1994SoPh..155..235M}Metcalf, T.R.: 1994, {\it Solar
Phys.} {\bf 155}, 235.
\bibitem[Metcalf \emph{et al.}(2006)]{2006SoPh..237..267M}Metcalf, T.R.,
Leka, K.D., Barnes, G., Lites, B.W., Georgoulis, M.K., Pevtsov, A.A., {\it et al.}
: 2006, {\it Solar Phys.}
{\bf 237}, 267.
\bibitem[Pesnell \emph{et al.}(2012)]{Pesnell2012}
Pesnell, W.D., Thompson, B.J., Chamberlin, P.C.: 2012, {\it Solar Phys.} {\bf 275}, 3.
\bibitem[Petrie and Patrikeeva(2009)]{2009ApJ...699..871P}Petrie,
G.J.D., Patrikeeva, I.: 2009, {\it Astrophys. J.} {\bf 699}, 871.
\bibitem[Shrauner and Scherrer(1994)]{1994SoPh..153..131S}Shrauner,
J.A., Scherrer, P.H.: 1994, {\it Solar Phys.} {\bf 153}, 131.
\bibitem[Svalgaard, Duvall, and
Scherrer(1978)]{1978SoPh...58..225S}Svalgaard, L., Duvall, T.L., Jr.,
Scherrer, P.H.: 1978, {\it Solar Phys.} {\bf 58}, 225.
\bibitem[Tsuneta \emph{et al.}(2008)]{2008SoPh..249..167T}Tsuneta, S.,
Ichimoto, K., Katsukawa, Y., Nagata, S., Otsubo, M., Shimizu, T., {\it et al.}: 2008, {\it Solar Phys.} {\bf 249}, 167.
\end{thebibliography}

\end{article}

\end{document}